\documentclass[twocolumn,floatfix,aps]{revtex4}
\usepackage{graphicx}
\usepackage{amsmath}
\usepackage{amssymb}
\usepackage{bm}

\begin{document}

\title{Dynamical signatures of ground-state degeneracy to discriminate against Andreev levels in a Majorana fusion experiment}
\author{A. Grabsch}
\affiliation{Instituut-Lorentz, Universiteit Leiden, P.O. Box 9506, 2300 RA Leiden, The Netherlands}
\author{Y. Cheipesh}
\affiliation{Instituut-Lorentz, Universiteit Leiden, P.O. Box 9506, 2300 RA Leiden, The Netherlands}
\author{C.W.J. Beenakker}
\affiliation{Instituut-Lorentz, Universiteit Leiden, P.O. Box 9506, 2300 RA Leiden, The Netherlands}

\date{September 2019}
\begin{abstract}
Detection of the fusion rule of Majorana zero-modes is a near-term milestone on the road to topological quantum computation. An obstacle is that the non-deterministic fusion outcome of topological zero-modes can be mimicked by the merging of non-topological Andreev levels. To distinguish these two scenarios, we search for dynamical signatures of the ground-state degeneracy that is the defining property of non-Abelian anyons. By adiabatically traversing parameter space along two different pathways one can identify ground-state degeneracies from the breakdown of adiabaticity. We show that the approach can discriminate against accidental degeneracies of Andreev levels.
\end{abstract} 
\maketitle

\section{Introduction}
\label{intro}

Non-Abelian anyons hold much potential for a quantum information processing that is robust to decoherence \cite{Kit97,Nay08}. The qubit degree of freedom is protected from local sources of decoherence since it is encoded nonlocally in a ground-state manifold of exponentially large degeneracy (of order $d^M$ for $M$ anyons with quantum dimension $d>1$). The degeneracy is called topological to distinguish it from accidental degeneracies that require fine tuning of parameters. The non-Abelian statistics follows from the ground-state degeneracy because exchange operations (braiding) correspond to non-commuting unitary operations in the ground-state manifold \cite{Wan15}.

Majorana zero-modes, midgap states in a superconductor, are non-Abelian anyons with quantum dimension $d=\sqrt 2$ \cite{Rea00,Das15}: Two zero-modes may or may not share an unpaired fermion, so that the ground state of $M$ zero-modes has degeneracy $2^{M/2}$.  To demonstrate the topological degeneracy of Majorana zero-modes is a near-term milestone on the road towards a quantum computer based on Majorana qubits \cite{Aas16}.

The general strategy for such a demonstration has been put forward by Aasen \textit{et al.} \cite{Aas16}. A set of four Majorana zero-modes $\gamma_1,\gamma_2,\gamma_3,\gamma_4$ is pairwise coupled (fused) in two different ways: Either $\gamma_2$ with $\gamma_3$ or $\gamma_1$ with $\gamma_2$. The zero-modes are then decoupled and the fermion parity $P_{12}$ of $\gamma_1$ and $\gamma_2$ is measured ($P_{12}=+1$ for even fermion number and $P_{12}=-1$ for odd fermion number). The $E=0$ ground-state degeneracy manifests itself in a nondeterministic outcome in the first case, with expectation value $\bar{P}_{12}=0$. The second case serves as a control experiment with a deterministic outcome of $+1$ or $-1$ depending on the sign of the coupling.

A challenge for the approach is formed by the tendency for non-topological Andreev levels to accumulate at $E=0$, resulting in a mid-gap peak in the density of states and a proliferation of accidental ground-state degeneracies \cite{Bee15}. The ground-state wave function of a few Andreev levels has local fermion-parity fluctuations that may mimic the non-deterministic fusion of Majorana zero-modes \cite{Cla17,Gra19}.

Here we present a dynamical description of the fusion strategy of Aasen \textit{et al.}, to search for signatures that make it possible to exclude spurious effects from Andreev levels. We traverse the parameter space of coupling constants along two pathways A and B such that the fermion parity measurement is non-deterministic along both pathways, but with identical expectation value $\bar{P}_{12}(\text{A})=\bar{P}_{12}(\text{B})$ when the evolution is adiabatic. Ground-state degeneracies are identified from the breakdown of adiabaticity, which causes $\bar{P}_{12}(\text{A})\neq \bar{P}_{12}(\text{B})$ in a way that is statistically distinct for Andreev levels and Majorana zero-modes.

\section{Adiabatic evolution to test for ground-state degeneracy}
\label{description}

\begin{figure}[tb]
\centerline{\includegraphics[width=1\linewidth]{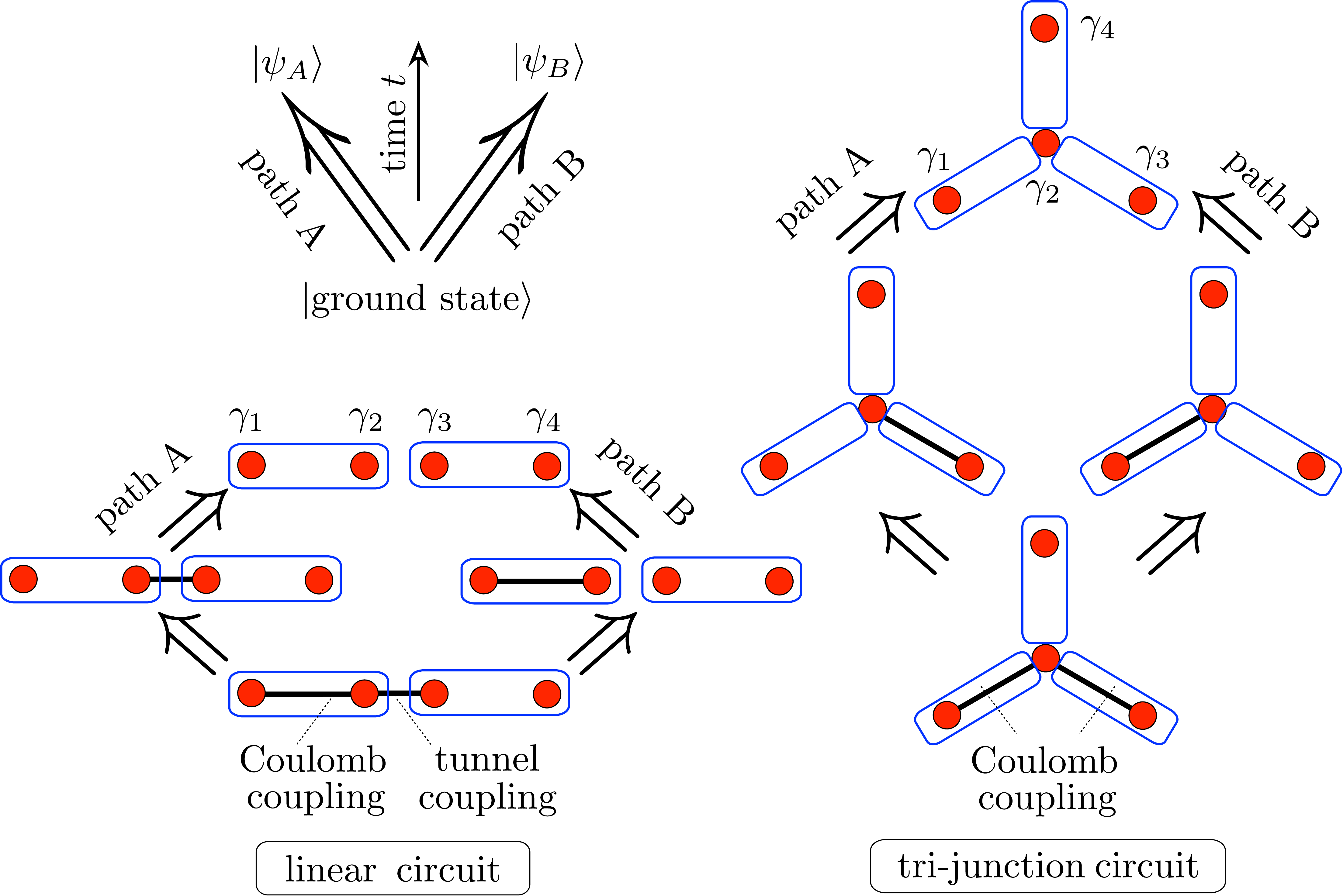}}
\caption{Two pathways A and B for the evolution of a Majorana qubit, encoded in four Majorana zero-modes (red dots) in a linear or tri-junction geometry. The blue contours represent superconducting islands and the black solid lines indicate which zero-modes are coupled. At the end of the evolution the Hamiltonian is the same for both pathways, but the final states $|\psi_{\text{A}}\rangle$ or $|\psi_{\text{B}}\rangle$ may depend on the pathway if adiabaticity breaks down because of a degenerate ground state.}
\label{fig_layout}
\end{figure}

We consider a Majorana qubit consisting of 4 Majorana zero-modes with 3 adjustable couplings, in either a linear geometry or a tri-junction geometry, see Fig.\ \ref{fig_layout}. The linear circuit contains two superconducting islands with adjustable Coulomb couplings in each island and a tunnel coupling between the islands. In the tri-junction there are three strongly coupled islands and only the Coulomb coupling within each island is adjustable.

The state $|\pm\rangle$ of the Majorana qubit is encoded in the fermion parity of one of the islands, say the island containing Majorana zero-modes 1 and 2. The fermion parity operator $P_{12}=-2i\gamma_1\gamma_2$ is the product of the two Majorana operators. Its eigenvalues are $+1$ or $-1$ depending on whether the fermion parity in that island is even or odd. For definiteness we will assume that the fermion parity of the entire system is even, and then $P_{34}=P_{12}$.

As illustrated in Fig.\ \ref{fig_layout}, in each geometry the system is initialized in the ground state with two of the three couplings on and the third coupling off. The final state with all couplings off is reached via one of two pathways, A or B, depending on which coupling is turned off first. 

Notice that at each instant in time the system contains at least two uncoupled zero-modes: $\gamma_4$ and an $E=0$ superposition $\gamma_0$ of $\gamma_1$, $\gamma_2$, $\gamma_3$ (which must exist because of the $\pm E$ symmetry of the spectrum \cite{Ali11}). Pathway A is the fusion process discussed by Aasen \textit{et al.} \cite{Aas16}, while pathway B is an element in the braiding process of Ref.\ \onlinecite{Hec12}.

If the ground state remains nondegenerate during this dynamical process, separated from excited states by a gap $E_{\rm gap}$ larger than the decoupling rate, then the adiabatic theorem ensures that the final state $|\psi\rangle_{\text{A}}=|\psi\rangle_{\text{B}}$ does not depend on the pathway. By measuring the expectation values
\begin{equation}
\bar{P}_{\text{A}}=\langle\psi_{\text{A}}|P_{12}|\psi_{\text{A}}\rangle,\;\;\bar{P}_{\text{B}}=\langle\psi_{\text{B}}|P_{12}|\psi_{\text{B}}\rangle,\label{PsiAPsiBdef}
\end{equation}
one can detect a breakdown from adiabaticity. This might be due to an accidental gap closing during the evolution, or due to the topological ground-state degeneracy of Majorana zero-modes.

We will consider the effect of an accidental degeneracy in Sec.\ \ref{accdeg}, in the next section we first address the topological degeneracy. 

\section{Topologically degenerate ground state}
\label{topdeg}

We summarize some basic facts about Majorana zero-modes (see reviews \cite{Nay08,Bee19} for more extensive discussions).

An even number $M=2N$ of uncoupled Majorana zero-modes has a $2^{N-1}$-fold degenerate ground-state manifold for a given global fermion parity. The degeneracy is removed by coupling, as described by the Hamiltonian
\begin{equation}
{H}=\tfrac{1}{2}\sum_{n,m=1}^{2N}{A}_{nm}i\gamma_n\gamma_m.\label{Hdef}
\end{equation}
The $2N\times 2N$ matrix ${A}$ is real antisymmetric, ${A}_{nm}=-{A}_{mn}={A}_{nm}^\ast$ and the Majorana operators $\gamma_n=\gamma_n^\dagger$ are Hermitian operators with anticommutator
\begin{equation}
\gamma_n\gamma_m+\gamma_m\gamma_n=\delta_{nm},\;\;\gamma_n^2=1/2.\label{Clifford}
\end{equation}

The fermion creation and annihilation operators $a^\dagger$, $a$ are related to the $\gamma$'s by
\begin{equation}
\begin{pmatrix}
\gamma_{2n-1}\\
\gamma_{2n}
\end{pmatrix}=U\begin{pmatrix}
a_{n}\\
a^\dagger_n
\end{pmatrix},\;\;U=\frac{1}{\sqrt 2}\begin{pmatrix}
1&1\\
-i&i
\end{pmatrix}.\label{gammaUdef}
\end{equation}
The fermion operators define a basis of occupation numbers, $|s_1,s_2,\ldots s_N\rangle$, such that $a^\dagger_{n}a_n|s_1,s_2,\ldots s_N\rangle=s_n|s_1,s_2,\ldots s_N\rangle$, $s_n\in\{0,1\}$.

For $N=2$ and assuming even global fermion parity the Hamiltonian \eqref{Hdef} in the basis of occupation numbers $|00\rangle\equiv|+\rangle$ and $|11\rangle\equiv|-\rangle$ reads
\begin{equation}
\begin{split}
&H=\tfrac{1}{2}\begin{pmatrix}
  -\Gamma & \Gamma'^\ast\\
\Gamma'&  \Gamma 
\end{pmatrix},\;\;\Gamma={A}_{12}+ {A}_{34},\\
&\Gamma'=-{A}_{14}-{A}_{23}-i {A}_{24} +i {A}_{13}.
\end{split}\label{HFock}
\end{equation}
The fermion parity operator ${P}_{12}=\sigma_z$ in this basis. Its expectation value in the ground state $|\text{GS}\rangle$ follows from
\begin{equation}
\begin{split}
&|\text{GS}\rangle\propto(\Gamma+\sqrt{\Gamma^2+|\Gamma'|^2})|+\rangle+\Gamma'|-\rangle\\
&\Rightarrow  \langle\text{GS}|P_{12}|\text{GS}\rangle=\frac{\Gamma}{\sqrt{\Gamma^2+|\Gamma'|^2}}.
\end{split}\label{GSdef}
\end{equation}

Eq.\ \eqref{GSdef} is a known result \cite{Cla17}, which shows that for $|\Gamma|\ll|\Gamma'|$ the ground state of the Majorana qubit is in an even-odd superposition of nearly equal weight. Applied to Fig.\ \ref{fig_layout} the same Eq.\ \eqref{GSdef} shows that the two pathways A and B correspond to an exchange of limits: $\Gamma\rightarrow 0$ before $\Gamma'\rightarrow 0$ for pathway A, resulting in $\bar{P}_{12}\rightarrow 0$, or the other way around for pathway B with $|\bar{P}_{12}|\rightarrow 1$.

\begin{figure}[tb]
\centerline{\includegraphics[width=0.7\linewidth]{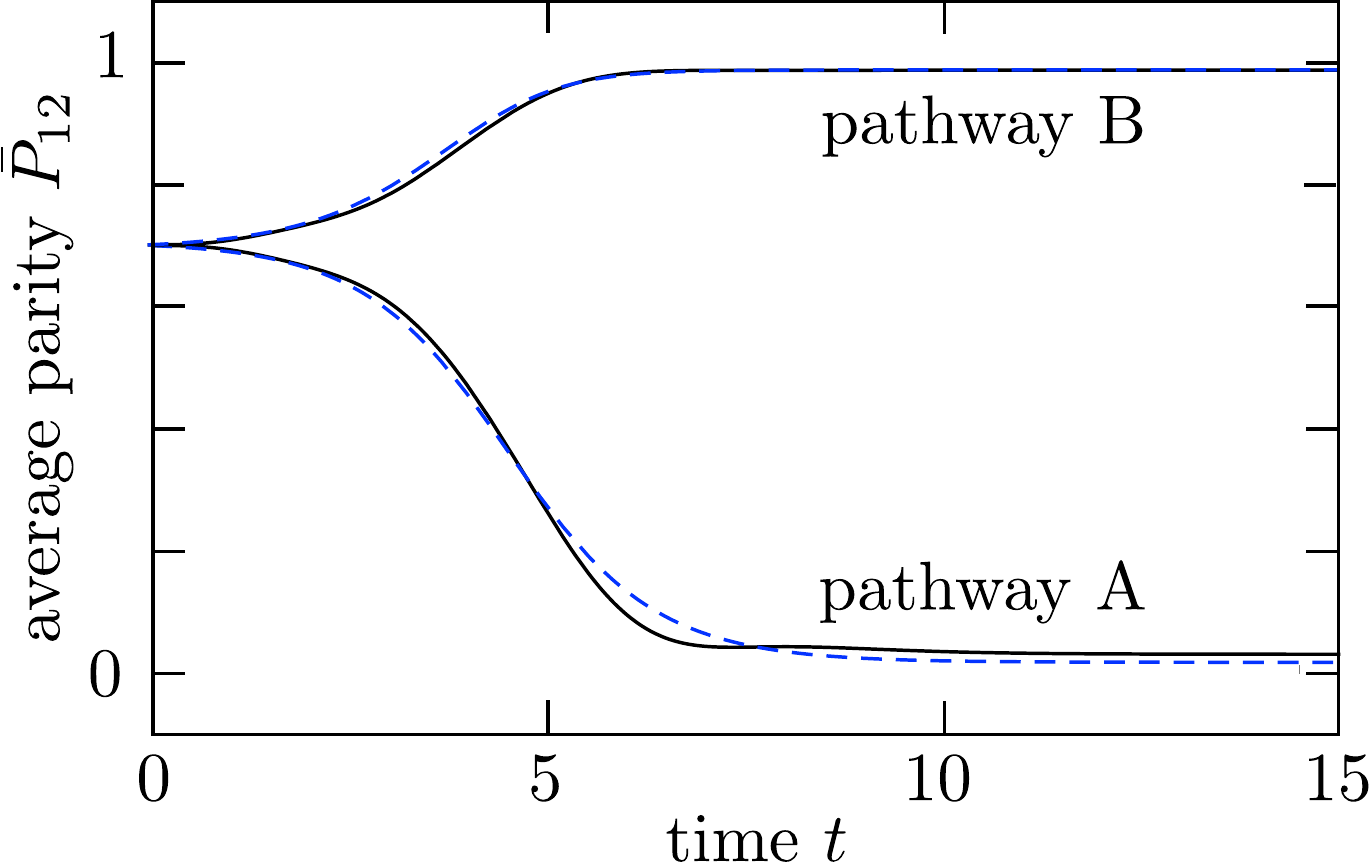}}
\caption{Solid curves: Expectation value $\bar{P}_{12}(t)=\langle\psi(t)|P_{12}|\psi(t)\rangle$ calculated numerically from the solution of the differential equation \eqref{dpsidt}, for the Hamiltonian \eqref{HFock} with time dependent coupling constants $\Gamma(t)=1-\tanh[(t-t_0)/\delta t]$ and $\Gamma'(t)=1-\tanh[(t-t'_0)/\delta t]$ for $\delta t =2$. The decoupling times are chosen at $t_0=4,t'_0=8$ for pathway A and $t_0=8,t'_0=4$ for pathway B. The dashed curves show the corresponding evolution of the expectation value in the ground state of $H(t)$, calculated from Eq.\ \eqref{GSdef}. The close agreement of solid and dashed curves indicates that the dynamics is nearly adiabatic.
}
\label{fig_dynamics}
\end{figure}

In Fig.\ \ref{fig_dynamics} we show how this works out dynamically, by integrating the evolution equation
\begin{equation}
i\hbar\frac{\partial}{\partial t}|\psi(t)\rangle=H\bigl(\Gamma(t),\Gamma'(t)\bigr)|\psi(t)\rangle,\label{dpsidt}
\end{equation}
with initial condition that $|\psi(0)\rangle$ is the ground state of $H$ at $t=0$.

\section{Accidentally degenerate Andreev levels}
\label{accdeg}

\begin{figure}[tb]
\centerline{\includegraphics[width=0.6\linewidth]{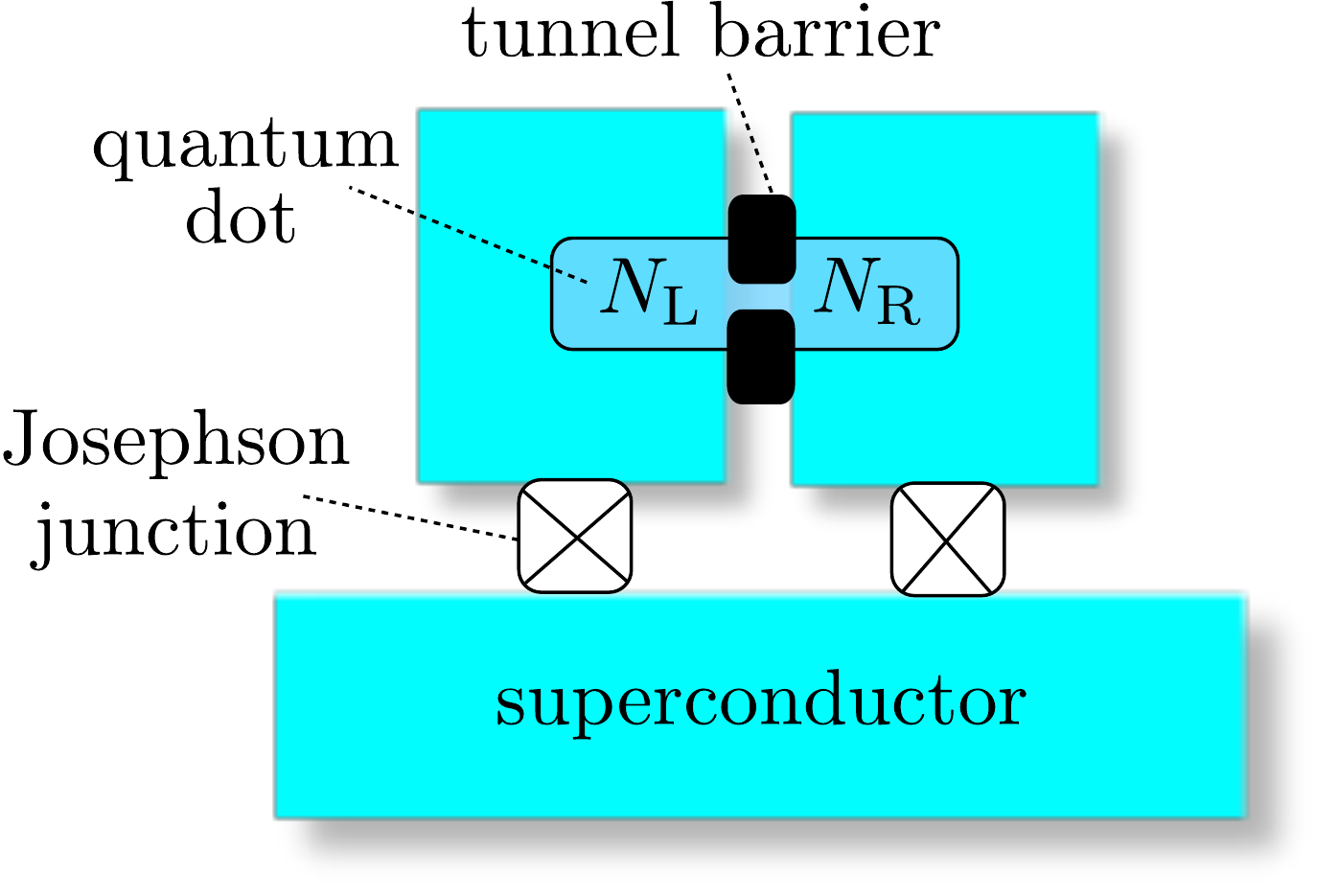}}
\caption{Two quantum dots on a superconducting substrate (blue), containing $N_{\rm L}$ and $N_{\rm R}$ Andreev levels coupled via a tunnel barrier. The coupling strength is adjustable via a pair of gate electrodes (black). The fermion parity $P_{\rm L}$, $P_{\rm R}$ in each quantum dot is regulated by the ratio $E_{\rm J}/E_{\rm C}$ of Josephson and charging energies, which is adjustable via the magnetic flux through a Josephson junction. In this way we can drive the system away from the ground state via the two pathways of Fig.\ \ref{fig_layout}, either by switching off first the fermion-parity coupling and then the tunnel coupling (pathway A) or the other way around (pathway B). At the end of each process the fermion parity $P_{\rm L}$ is measured.}
\label{fig_doubledot}
\end{figure}

To assess the breakdown of the adiabatic evolution as a result of (nearly) degenerate Andreev levels we consider the double quantum dot geometry of Fig.\ \ref{fig_doubledot}. There are $N_{\rm L}$ Andreev levels in the left dot and $N_{\rm R}$ Andreev levels in the right dot. The quantum dots are coupled to each other by an adjustable tunnel barrier and each has an adjustable coupling to a bulk superconductor by a Josephson junction. 

For strong Josephson coupling the Coulomb charging energy may be neglected and the Hamiltonian of the double-quantum dot is bilinear in the creation and annihilation operators,
\begin{subequations}
\label{HBdef}
\begin{align}
&{\cal H}_0=\tfrac{1}{2}\sum_{n,m=1}^{N}\Psi_n^\dagger\cdot{\cal B}_{nm}\cdot\Psi_m,\label{HBdefa}\\
&\Psi_n=\begin{pmatrix}
a_n \\
a_n^\dagger
\end{pmatrix},\;\;{\cal B}_{nm}=\begin{pmatrix}
V_{nm}&-\Delta^\ast_{nm}\\
\Delta_{nm}&-V_{nm}^\ast
\end{pmatrix}.\label{HBdefb}
\end{align}
\end{subequations}
The indices $n,m$ label spin and orbital degrees of freedom of the $N=N_{\rm L}+N_{\rm R}$ Andreev levels. The $N\times N$ Hermitian matrix $V$ represents the kinetic and potential energy. The $N\times N$ antisymmetric matrix $\Delta$ is the pair potential.

As the ratio $E_{\rm J}/E_{\rm C}$ of Josephson and charging energy is reduced, the Coulomb interaction in a quantum dot becomes effective. In the regime $E_{\rm J}/E_{\rm C}\gtrsim 1$  the interaction term only depends on the fermion parity \cite{Hec12},
\begin{equation}
\begin{split}
&{\cal H}_{\rm C}=-\kappa_{\rm L}P_{\rm L}-\kappa_{\rm R}P_{\rm R},\\
&P_{\rm L}=(-1)^{\sum_{n\in \text{L}}a^\dagger_n a_n},\;\;
P_{\rm R}=(-1)^{\sum_{n\in \text{R}}a^\dagger_n a_n}.
\end{split}\label{HCdef}
\end{equation}
The two coupling constants $\kappa_{\rm L}$ and $\kappa_{\rm R}$ depend exponentially $\propto e^{-\sqrt{8E_{\rm J}/E_{\rm C}}}$ on the Josephson energy  \cite{Mak01}, which can be varied by adjusting the magnetic flux through the Josephson junction connected to the left or right quantum dot. We set $\kappa_{\rm R}\equiv 0$ for all times while $\kappa_{\rm L}(t)$ drops from $\kappa_0$ to $0$ in an interval $\delta t$ around $t=t_0$. We choose a tanh profile,
\begin{equation}
\kappa_{\rm L}(t)=\tfrac{1}{2} \kappa_0-\tfrac{1}{2} \kappa_0\tanh[(t-t_0)/\delta t].\label{kappaLt}
\end{equation}

\begin{figure*}[tb]
\centerline{\includegraphics[width=0.9\linewidth]{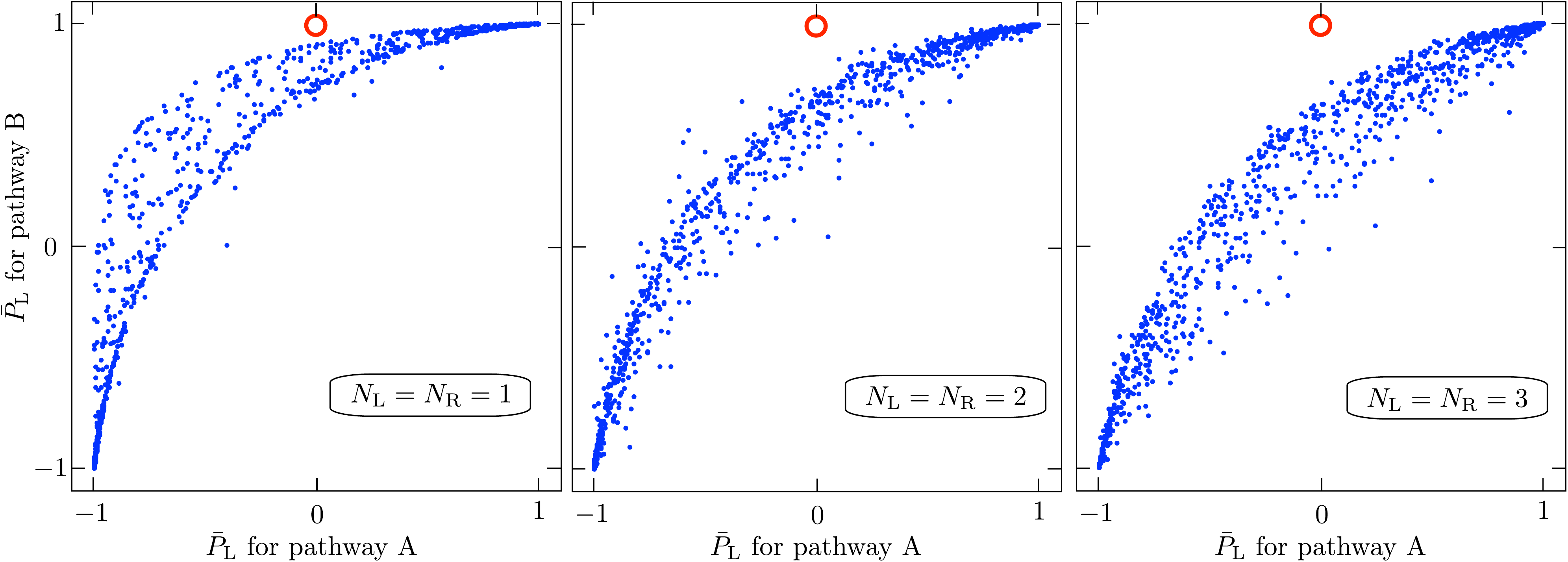}}
\caption{Scatter plot that illustrates how the expectation value $\bar{P}_{\rm L}$ of the fermion parity in the left quantum dot depends on the pathway A or B that is followed in parameter space. Each blue dot results from one realization of the class-D ensemble of random Hamiltonians ${\cal H}_0$. In units such that the mean Andreev level spacing $\delta_0\equiv 1$, the parameters in Eqs.\ \eqref{kappaLt} and \eqref{kappaLRdef} are $\delta t=\delta t'=2$, $\kappa_0=1/4$ for both pathways, and $t_0=4,t'_0=8$ for pathway A, $t_0=8,t'_0=4$ for pathway B. The fermion parity is evaluated at time $t=15$. The red circle indicates the expected outcome for a Majorana qubit, which is well separated from the scatter plot of Andreev levels.
}
\label{fig_scatterplot}
\end{figure*}

For each of the two dynamical pathways A and B we start at $t=0$ with a strong tunnel coupling between the quantum dots. We model this statistically by means of the Gaussian ensemble of random-matrix theory in symmetry class D (broken time-reversal and broken spin-rotation symmetry) \cite{Bee15,Alt97}.

The ensemble is constructed as follows. A unitary transformation to the Majorana basis,
\begin{equation}
U {\cal B}_{nm} U^\dagger=i{\cal A}_{nm},\;\;U=\frac{1}{\sqrt 2}\begin{pmatrix}
1&1\\
-i&i
\end{pmatrix},\label{ABrelation}
\end{equation}
see Eq.\ \eqref{gammaUdef}, expresses the Hamiltonian \eqref{HBdef} in terms of a real antisymmetric $2N\times 2N$ matrix ${\cal A}$. We take independent Gaussian distributions for each upper-diagonal matrix element of ${\cal A}$, with zero mean and variance  $2N\delta_0^2/\pi^2$, where $\delta_0$ is the mean spacing of the Andreev levels.
 
For strongly coupled quantum dots we do not distinguish statistically between matrix elements ${\cal A}_{nm}$ that refer to levels $n$ and $m$ in the same dot or in different dots. To decouple the quantum dots by the tunnel barrier we suppress the inter-dot matrix elements,
\begin{subequations}
\label{kappaLRdef}
\begin{align}
&{\cal A}_{nm}(t)={\cal A}_{nm}(0)\times\begin{cases}
1&\text{if}\;n,m\;\text{in the same dot},\\
\kappa_{\rm LR}(t)&\text{if}\;n,m\;\text{in different dots},
\end{cases}\\
&\kappa_{\rm LR}(t)=\tfrac{1}{2} -\tfrac{1}{2} \tanh[(t-t'_0)/\delta t'].\label{kappaprimetdef}
\end{align}
\end{subequations}

We solve the Schr\"{o}dinger equation
\begin{equation}
i\hbar\frac{\partial}{\partial t}|\psi\rangle=({\cal H}_0+{\cal H}_{\rm C})|\psi\rangle,\label{dpsidtH0HC}
\end{equation}
by first calculating the Hamiltonian in the $2^{N_{\rm L}+N_{\rm R}-1}$ dimensional basis of occupation numbers in the left and right dot, for even global fermion parity $P_{\rm L}P_{\rm R}=+1$. (We used the {\sc sneg} package to take over this tedious calculation \cite{sneg}.) Starting from the ground state at $t=0$ we switch off $\kappa_{\rm L}$ and $\kappa_{\rm LR}$ along pathways A or B (first switching off $\kappa_{\rm L}$ or first switching off $\kappa_{\rm LR}$, respectively). At the end of the process we calculate the expectation value of the fermion parity $\bar{P}_{\rm L}$ in the left dot. 

The calculation is repeated for a large number of realizations of the Hamiltonian ${\cal H}_0$ in the class-D ensemble. A scatter plot of $\bar{P}_{\rm L}({\rm A})$ versus $\bar{P}_{\rm L}({\rm B})$ is shown in Fig.\ \ref{fig_scatterplot} for a few values of $N_{\rm L},N_{\rm R}$. Significant deviations are observed from the line $\bar{P}_{\rm L}({\rm A})=\bar{P}_{\rm L}({\rm B})$ of adiabatic evolution, but the scatter plot stays clear of the point $\bar{P}_{\rm L}({\rm A})=0$, $\bar{P}_{\rm L}({\rm B})=1$ that characterizes a Majorana qubit.

\begin{figure}[tb]
\centerline{\includegraphics[width=0.7\linewidth]{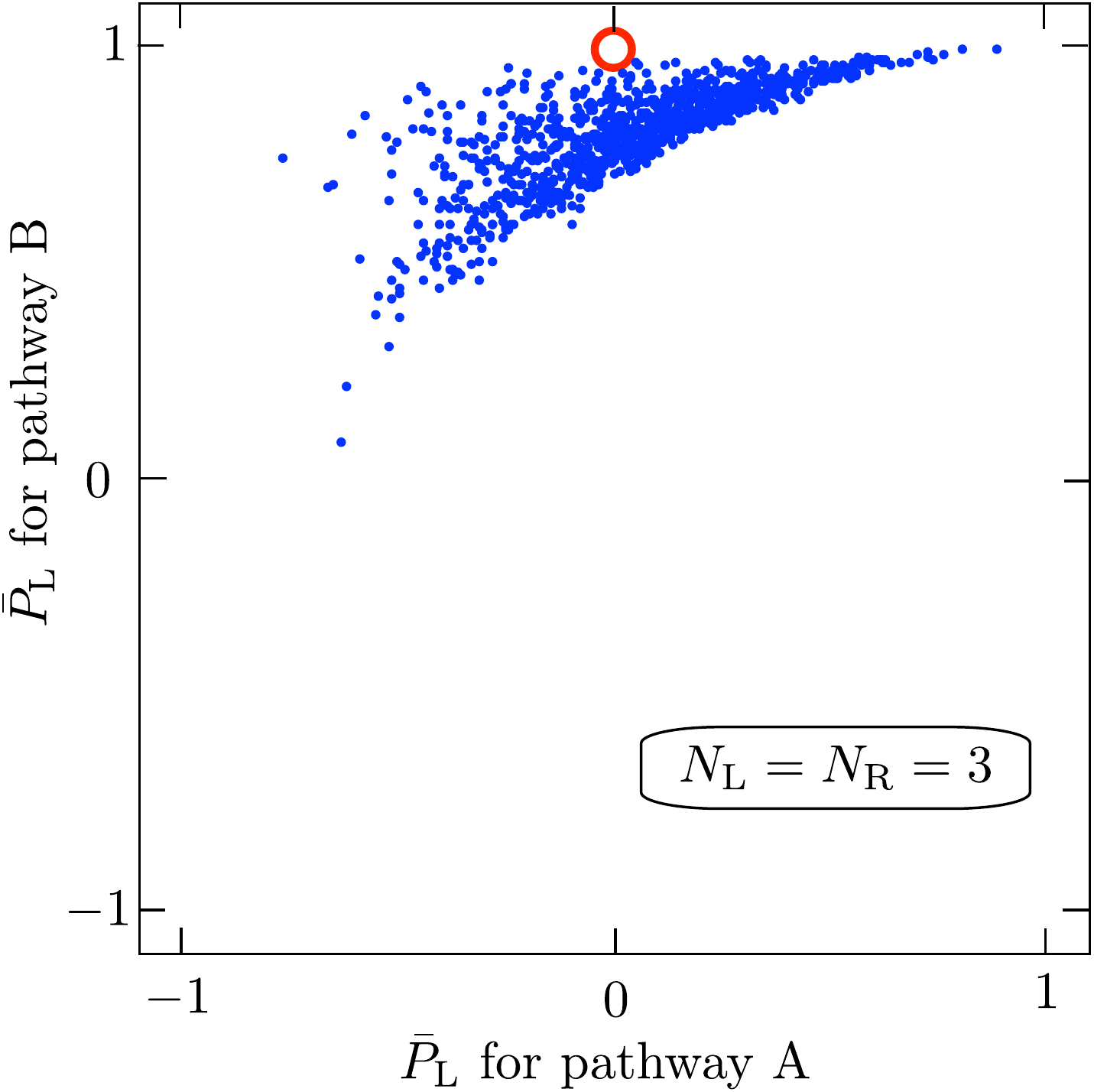}}
\caption{Same as Fig.\ \ref{fig_scatterplot}, but now for a stronger fermion-parity coupling ($\kappa_0=2$) and abrupt removal of the tunnel coupling ($\delta t'=1/4$, all other parameters unchanged). The outcome for a Majorana qubit is now no longer well separated from the scatter plot of the outcome from Andreev levels.
}
\label{fig_scatterplot2}
\end{figure}

Two ingredients in the fusion protocol are essential for this to work: Firstly, the fermion-parity coupling should be smaller than or comparable to the tunnel coupling, in order for pathway B to have a nondeterministic fusion outcome. Secondly, the tunnel coupling should be cut slowly on the scale of the inverse mean level spacing, to promote adiabatic evolution in pathway A. In Fig.\ \ref{fig_scatterplot2} we show the scatter plot when both these conditions are violated: There is now no clear separation from the Majorana qubit.

\section{Conclusion}
\label{conclude}

A succesful demonstration of the non-deterministic fusion of two Majorana zero-modes would be a milestone in the development of a topological quantum computer \cite{Aas16}. Its significance would be both conceptual (because it implies non-Abelian braiding statistics \cite{Wan15}) and practical (because fusion can substitute for braiding in a quantum computation \cite{Bee19,Lit17}).

In this work we have investigated the dynamics of the fusion process, to see how spurious effects from the merging of Andreev levels can be eliminated. We compare the time-dependent evolution in the parameter space of coupling constants (tunnel coupling and Coulomb coupling) via two alternative pathways. The topological ground-state degeneracy of Majorana zero-modes causes a breakdown of adiabaticity that can be measured as a pathway-dependent fermion parity. Andreev levels can produce accidental degeneracies, and a non-deterministic fermion parity outcome, but the correlation between the two pathways is distinct from what would follow from the Majorana fusion rule (see Fig.\ \ref{fig_scatterplot}).

Initial experimental steps towards the detection of the Majorana fusion rule have been reported \cite{Raz17}. Typical spacings $\delta_0$ of sub-gap Andreev levels in these nanowire geometries are $ 10\,\mu{\rm eV}$, so the adiabatic decoupling time scale $\delta t=2\hbar/\delta_0$ in Fig.\ \ref{fig_scatterplot} would be on the order of $0.1\,{\rm ns}$, well below expected quasiparticle poisoning times of $1\,\mu{\rm s}$ \cite{Alb17}.
 
\acknowledgments

This project has received funding from the Netherlands Organization for Scientific Research (NWO/OCW) and from the European Research Council (ERC) under the European Union's Horizon 2020 research and innovation programme.

\end{document}